\documentclass[12pt]{iopart}
\usepackage{iopams}
\usepackage{graphicx}

\newcommand{\Rm}{\mathbb{R}}

\newcommand{\be}{\begin{equation}}
\newcommand{\ee}{\end{equation}}
\newcommand{\ba}{\begin{eqnarray}}
\newcommand{\ea}{\end{eqnarray}}

\newcommand{\pp}{\partial}

\newcommand{\argmin}{\mathop{\mathrm{arg\,min}}}

\begin{document}

\title[]{Shape-constrained reconstruction in diffuse optical tomography by simulated annealing}

\author{Tetsuya Mimura$^{1,2}$, Yu Jiang$^3$, Norikazu Todoroki$^4$, and Manabu Machida$^1$}
\address{${^1}$ Institute for Medical Photonics Research, Hamamatsu University School of Medicine, Hamamatsu 431-3192, Japan}
\address{${^2}$ Central Research Laboratory, Hamamatsu Photonics K.K., Hamamatsu, 434-8601, Japan}
\address{${^3}$ School of Mathematics, Shanghai University of Finance and Economics, Shanghai 200433, P.R. China}
\address{${^4}$ Department of Physics, Chiba Institute of Technology, Chiba 275-0023, Japan}


\begin{abstract}
When the inverse problem of diffuse optical tomography (DOT) is solved with the Born or Rytov approximation, the size of the matrix of the linear inverse problem becomes large if the volume (or area) of the domain in biological tissue used for reconstruction is large. The number of unknown parameters in DOT is reduced when constraints about the shape of a target are imposed for the inverse problem. Due to such constraints, the inverse problem becomes nonlinear even when the (first) Born or Rytov approximation is employed. We solve this nonlinear inverse problem by the simulated annealing, which is not trapped by local minima of the cost function.
\end{abstract}

%
\vspace{2pc}

\section{Introduction}

Diffuse optical tomography (DOT), which is one of medical imaging modalities, uses near-infrared light. It is known that the inverse problem of DOT is severely ill-posed \cite{Isakov06}. Hence the resolution of tomographic images of DOT is limited. In this paper, the reconstruction of detailed structures of the target is not attempted but we impose shape constraints when solving the inverse problem. Since the shape of a target is a priori assumed, the number of unknowns can be significantly reduced.

The diffusion coefficient, absorption coefficient, or both in the diffusion equation are reconstructed in DOT \cite{Arridge99}. The choice of good initial guesses is essential when these nonlinear inverse problems are solved by iterative schemes such as the conjugate gradient method and Gauss-Newton method \cite{Dennis-Schnabel83,Nocedal-Wright06}. As an alternative approach, the Born and Rytov series are often employed in DOT \cite{Arridge-Schotland09}. Usually they are used to linearize nonlinear inverse problems with the (first-order) Born or Rytov approximation. The issue of initial guesses can be avoided by such direct methods. Instead, optical properties of the reference medium are necessary for the direct methods. Examples of the use of the Rytov approximation include an experiment of the optical tomography with structured illumination \cite{Konecky-etal09} and functional DOT \cite{Doulgerakis-Eggebrecht-Dehghania19}.

When the shape of the target to be reconstructed is fixed, the relation between the solution of the diffusion equation and the shape parameters becomes nonlinear even when the original inverse problem between the solution and coefficients of the diffusion equation is linearized by the Born or Rytov approximation. One way of solving this nonlinear inverse problem is to rely on iterative methods. Then, however, the issue of the choice of initial guesses arises again. In \cite{Sun-etal20}, the linear inverse problem of fluorescence diffuse optical tomography was considered. When the shape of a target is assumed to be a cuboid, the inverse problem becomes nonlinear. In \cite{Sun-etal20}, the choice of good initial guesses was important to identify the target size and position by the Levenberg–Marquardt algorithm.

Compared with iterative methods, Monte Carlo methods can reach the global minimum of the cost function without trapping by local minima. The pilot adaptive Metropolis algorithm was employed for the electrical impedance tomography \cite{Ahmad-etal19}. In general, Bayesian inverse schemes do not converge or very slowly converge when there are many unknown parameters \cite{Lassas-Siltanen04,Lucka12}. Although the use of Monte Carlo methods has been attempted in studies related to DOT, their computations were time-consuming \cite{Bal-etal13,Barnett-etal03,Langmore13}.

In this paper, the Monte Carlo approach is employed. The computation is brought to converged values by simulated annealing. We will solve the inverse problem of DOT by fixing the shape of the spatial distribution of the absorption coefficient in the diffusion equation, while the true shape of the target is not necessarily an assumed shape. By this, the computation time is significantly reduced. It is shown that the position of the inhomogeneity of the absorption coefficient is identified by our method using numerical phantoms in two and three dimensions.

The remainder of the paper is organized as follows. In Sec.~\ref{method}, we formulate our algorithm for DOT and explain settings of two- and three-dimensional numerical phantoms. Results of our numerical experiments are illustrated in Sec.~\ref{results}. Secs.~\ref{discussion} and \ref{concl} are devoted to discussion and conclusion, respectively. In \ref{green}, the Green's function for the two-dimensional diffusion equation is given.

\section{Method}
\label{method}

\subsection{Diffusion equation}
\label{DE}

Let us consider diffuse light in a domain $\Omega\subset\Rm^d$ ($d=2,3$). Let $\pp\Omega$ be the boundary of $\Omega$. Let $\nu(x)$ be the unit outer normal vector at $x\in\pp\Omega$. We assume that $\Omega$ is occupied by biological tissue and $\Rm^d\setminus\overline{\Omega}$ is vacuum. The domain $\Omega$ is characterized by absorption coefficient $\mu_a(x)$ and diffusion coefficient $D_0$. We assume that $\mu_a(x)$ varies in space but $D_0$ is a positive constant. The diffuse fluence rate $u$ obeys the following diffusion equation.
\begin{equation}
\cases{
-D_0\Delta u+\mu_au=f,
&$x\in\Omega$,
\\
D_0\pp_{\nu}u+\frac{1}{\zeta}u=0,
&$x\in\pp\Omega$,
}
\label{de0}
\end{equation}
where $\zeta$ is a positive constant. We assume that $\mu_a\in L^{\infty}(\Omega)$, $\mu_a>0$, and $\left.\mu_a\right|_{\pp\Omega}=\mu_{a,0}$ with $\mu_{a,0}$ a positive constant. The constant $\zeta$ is determined from the reflection on the boundary. We assume the diffuse surface reflection and give $\zeta$ by \cite{Egan-Hilgeman79}
\be
\zeta=2\frac{1+r_d}{1-r_d},\quad
r_d=-1.4399\mathfrak{n}^{-2}+0.7099\mathfrak{n}^{-1}+0.6681+0.0636\mathfrak{n}
\ee
with the ratio $\mathfrak{n}$ of refractive indices inside and outside the medium $\Omega$. Near-infrared light is illuminated at a point and the outgoing light is detected at another point on the boundary. We suppose there are $M_{\rm SD}$ source-detector pairs. The incident beam $f(x)$ is assumed to be
\be
f(x)=g_0\delta(x-x_s^{(l)}),
\ee
where $g_0>0$ is a constant, $x_s^{(l)}$ is the position of the source of the $l$th source-detector pair ($l=1,2,\dots,M_{\rm SD}$), and $\delta(\cdot)$ is the Dirac delta function. The absorption coefficient $\mu_a(x)$ can  be expressed as
\be
\mu_a(x)=\mu_{a,0}\left(1+\eta(x)\right),\quad x\in\Omega.
\ee

Although $\eta(x)$ may be a complicated function of $x$, we will reconstruct $\eta$ by fixing the shape of $\eta$. Let $\eta_*$ be the reconstructed $\eta$. Let $\Omega_{\rm recon}\subset\Omega$ be the domain of the assumed fixed-shape target $\eta_*$. We assume that $\mu_a$ takes a constant value $\Omega_{\rm recon}$. That is, $\eta_*=\eta_0$ ($\eta_0>0$ is a constant) at the target and $\eta_*=0$ in the background:
\be
\eta_*(x)=\cases{
\eta_0,&$x\in\Omega_{\rm recon}$,
\\
0,&$x\in\Omega\setminus\Omega_{\rm recon}$.
}
\ee
Thus, instead of reconstructing $\mu_a(x)$ in $\Omega$, we try to find $\Omega_{\rm recon}$ and $\eta_0$. Let $N$ be the number of parameters that are necessary to determine $\Omega_{\rm recon}$ and $\eta_0$. The shape-constrained reconstruction has been developed in fluorescence DOT \cite{Sun-etal20,Wang-Liu20}. In this paper, we will incorporate this tomography in the simulated annealing for DOT.

By taking a sufficiently large $\eta_0^{\rm(max)}$, we can restrict $\eta_0$ in the range $(-1,\eta_0^{\rm(max)}]$.  Let $M$ be a positive integer. We introduce
\be
S_i=0,\pm1,\pm2,\dots,\pm M\quad(i=1,\dots,N).
\ee
We will determine $N$ unknown parameters using $S_i$ ($i=1,\dots,N$). 
Let $m$ be an integer which is defined as
\be
m=\min\left(\left\lfloor\frac{M}{\eta_0^{\rm(max)}}\right\rfloor,M\right).
\ee
Using $S_N$, we give $\eta_0$ as
\be
\eta_0=\frac{\eta_0^{\rm(max)}}{M}S_N,\quad-m\le S_N\le M.
\ee
Other spins $S_1$ through $S_{N-1}$ will be described in Sec.~\ref{num2d}. The solution $u(x)=u[S](x)$ of (\ref{de0}) is uniquely obtained for each configuration of $S=(S_1,\dots,S_N)$.

\subsection{Measurement}
\label{rytov}

As a reference we consider the following diffusion equation with $\mu_{a,0}$.
\begin{equation}
\cases{
-D_0\Delta u_{\rm ref}+\mu_{a,0}u_{\rm ref}=f,
&$x\in\Omega$,
\\
D_0\pp_{\nu}u_{\rm ref}+\frac{1}{\zeta}u_{\rm ref}=0,
&$x\in\pp\Omega$.
}
\label{de1}
\end{equation}
Indeed, $u_{\rm ref}(x)$ is the Green's function $G(x,x_s^{(l)})$ with the relation
\be
u_{\rm ref}(x)=g_0G(x,x_s^{(l)}).
\ee

Suppose that light is detected at $x_d^{(l)}\in\pp\Omega$ for the $l$th source-detector pair. We consider the following data $\phi^{(l)}$.
\begin{equation}
\phi^{(l)}(S)=\ln\frac{u_{\rm ref}(x_d^{(l)})}{u(x_d^{(l)})}.
\label{datafunc}
\end{equation}
With the Rytov approximation, $\phi^{(l)}$ is given by
\begin{equation}
\phi^{(l)}(S)=\frac{\mu_{a,0}}{G(x_d^{(l)},x_s^{(l)})}
\int_{\Omega}G(x_d^{(l)},y)\eta(y)G(y,x_s^{(l)})\,dy.
\label{dataRytov0}
\end{equation}
By discretization we have
\begin{equation}
\phi^{(l)}(S)\approx
\frac{\mu_{a,0}\eta_0(\Delta y)^d}{G(x_d^{(l)},x_s^{(l)})}
\sum_{y_i\in\Omega_{\rm recon}}G(x_d^{(l)},y_i)G(y_i,x_s^{(l)}),
\label{dataRytov}
\end{equation}
where $y_i$ is the position of the representative point of the $i$th voxel in $\Omega_{\rm recon}$ and $(\Delta y)^d$ is the $d$-dimensional volume of a voxel. The corresponding measured data will be denoted by $\Phi^{(l)}$. 

\subsection{Simulated annealing}
\label{SA}

We solve the inverse problem by minimizing the following cost function $\mathcal{H}(S)$.
\begin{equation}
\mathcal{H}(S)=\frac{1}{2}\sum_{l=1}^{M_{\rm SD}}\left|\Phi^{(l)}-\phi^{(l)}(S)\right|^2+\varepsilon\|S-\bar{S}\|_{\ell^1},
\label{SA:cost}
\end{equation}
where $\bar{S}$ is the initial guess and
\be
\|S-\bar{S}\|_{\ell^1}=\sum_{i=1}^N|S_i-\bar{S}_i|.
\ee
The regularization parameter $\varepsilon$ is nonnegative. We wish to find the configuration $S_*=\argmin_{S}\mathcal{H}(S)$.

To solve the inverse problem with the simulated annealing, we introduce temperature $T$. The simulated annealing finds $S_*$ by decreasing temperature from $T_{\rm high}$ to $T_{\rm low}$. Let $f_{\rm prior}(S)$ be the prior distribution that is zero if any of $S_i$ ($i=1,\dots,N$) is outside the given interval and is a positive constant otherwise. The partition function $Z$ is given by
\be
Z=\sum_{\{S_i\}}e^{-\beta\mathcal{H}(S)}f_{\rm prior}(S),
\ee
where $\beta=1/T$ is the inverse temperature. Here, we used the notation $\sum_{\{S_i\}}=\sum_{S_1=-M}^M\cdots\sum_{S_N=-M}^M$. The probability density function $\pi(S)$ is given by
\be
\pi(S)=\frac{e^{-\beta\mathcal{H}(S)}f_{\rm prior}(S)}{Z}.
\label{probS}
\ee

The proposal distribution $q(S'_i|S_i)$ is given by generating the value $S'_i$ between $-M$ and $M$ at the $i$th site with equal probability. For two configurations $S,S'$, we have
\be
\frac{\pi(S')}{\pi(S)}=e^{\beta(\mathcal{H}(S)-\mathcal{H}(S'))}\quad
\mbox{when}\quad f_{\rm prior}(S)\neq0,\quad f_{\rm prior}(S')\neq0.
\ee
The acceptance probability is introduced as
\be
\alpha(S'_i,S_i)=
\min\left\{1,\frac{\pi(S_1,\dots,S'_i,\dots,S_N)}{\pi(S_1,\dots,S_i,\dots,S_N)}\right\},\quad i=1,\dots,N.
\ee
The transition kernel is given by
\be
K(S'_i,S_i)=
\alpha(S'_i,S_i)q(S'_i|S_i)+
\delta_{S'_i,S_i}\sum_{S'_i}\left(1-\alpha(S'_i,S_i)\right)q(S'_i|S_i).
\ee
We have $K(S'_i,S_i)\ge0$ and $\sum_{S'_i}K(S'_i,S_i)=1$. We note that the detailed balance below is satisfied for each pair $(S'_i,S_i)$.
\be
K(S'_i,S_i)\pi(S_1,\dots,S_i,\dots,S_N)=
K(S_i,S'_i)\pi(S_1,\dots,S'_i,\dots,S_N).
\ee
This is a necessary condition for $S\to S_*$.

Now we can perform the simulated annealing as follows.

\vspace{1em}
\textbf{Simulated annealing}
\begin{itemize}
\setlength\itemsep{1em}
\item[Step 1.]
Start with a small $\beta=1/T_{\rm high}>0$. Give initial $S_i$ ($i=1,\dots,N$) randomly. Then set $i=1$.
\item[Step 2.]
Generate $S'_i\sim q(S'_i|S_i)$.
\item[Step 3.]
Calculate $\alpha(S'_i,S_i)$.
\item[Step 4.]
Replace $S_i$ by $S'_i$ with probability $\alpha(S'_i,S_i)$.
\item[Step 5.]
Set $i=1$ if $i=N$. Otherwise set $i=i+1$. Return to Step 2. After arriving at the burn-in time, stop iterating the loops from Step 2 to Step 5 and proceed to Step 6.
\item[Step 6.]
Decrease temperature and go to Step 2. If the temperature reaches $T_{\rm low}$, finish the iteration.
\end{itemize}
\vspace{1em}

In this paper, we decrease temperature as
\be
T-10^{\mathop{\mathrm{int}}(\log_{10}T)-2}\quad\rightarrow\quad T.
\ee
At Step 5, the computation is run for $10$ Monte Carlo steps before moving to Step 6.

\subsection{Numerical experiments in two dimensions}
\label{num2d}

We consider the half space: $\Omega=\{x\in\Rm^2;\;-\infty<x_1<\infty,\;0<x_2<\infty\}$. We use $16$ sources and $15$ detectors, which results in $M_{\rm SD}=240$ source-detector pairs:
\be
x_{s1}^{(p)}=\pm2,\pm6,\dots,\pm30\,{\rm mm},\quad
x_{d1}^{(p)}=0,\pm4,\pm8,\dots,\pm28\,{\rm mm}.
\ee
We obtain the forward data by solving the diffusion equation with the finite-difference scheme. We added $3\%$ Gaussian noise to the forward data $\Phi^{(p)}$ ($p=1,\dots,M_{\rm SD}$). For the inverse problem, the Green's function is computed according to \ref{green}. We set
\be
\mu_{a,0}=0.02\,{\rm mm}^{-1},\quad D_0=0.33\,{\rm mm}.
\ee
Moreover the refractive index is set to $\mathfrak{n}=1.37$. A disk-shaped target of diameter $5\,{\rm mm}$ is placed in the medium. The center of the disk is at $(0,10\,{\rm mm})$. Inside the disk,
\be
\mu_a(x)=0.22\,{\rm mm}^{-1}.
\ee
This means $\eta(x)=10$ inside the disk.

We consider two kinds of $\Omega_{\rm recon}$: $\Omega_{\rm square}$ and $\Omega_{\rm disk}$. They are defined as
\ba
\Omega_{\rm square}
&=
\left\{x\in\Omega;\quad
\xi_1-\frac{\ell}{2}<x_1<\xi_1+\frac{\ell}{2},\quad
\xi_2-\frac{\ell}{2}<x_2<\xi_2+\frac{\ell}{2}\right\},
\nonumber \\
\Omega_{\rm disk}
&=
\left\{x\in\Omega;\quad
(x_1-\xi_1)^2+(x_2-\xi_2)^2<\left(\frac{\ell}{2}\right)^2\right\}.
\ea
Here, $\xi_1,\xi_2$ ($\xi=(\xi_1,\xi_2)\in\Omega$) and $\ell>0$ are unknown parameters to be determined. This means
\be
N=4.
\ee
We set constants $\xi_1^{\rm(max)},\xi_2^{\rm(max)}$,$\ell^{\rm(max)}$ such that $\xi_1$ is a constant in $[-\xi_1^{\rm(max)},\xi_1^{\rm(max)}]$, $\xi_2$ is a constant in $(0,\xi_2^{\rm(max)}]$, and $\ell$ is a constant in $(0,\ell^{\rm(max)}]$. We set
\ba
\xi_1&=
\frac{\xi_1^{\rm(max)}}{M}S_1,\quad-M\le S_1\le M,
\nonumber \\
\xi_2&= 
\frac{\xi_2^{\rm(max)}}{M}S_2,\quad 1\le S_2\le M,
\nonumber \\
\ell&=
\frac{\ell^{\rm(max)}}{M}S_3,\quad 1\le S_3\le M.
\ea
Finally, $f_{\rm prior}=0$ if $S_2$ or $S_3$ is not positive, or $S_4$ is less than $-m$,

\subsection{Numerical experiments in three dimensions}
\label{num3d}

\begin{figure}[ht]
\centering
\includegraphics[width=0.6\textwidth]{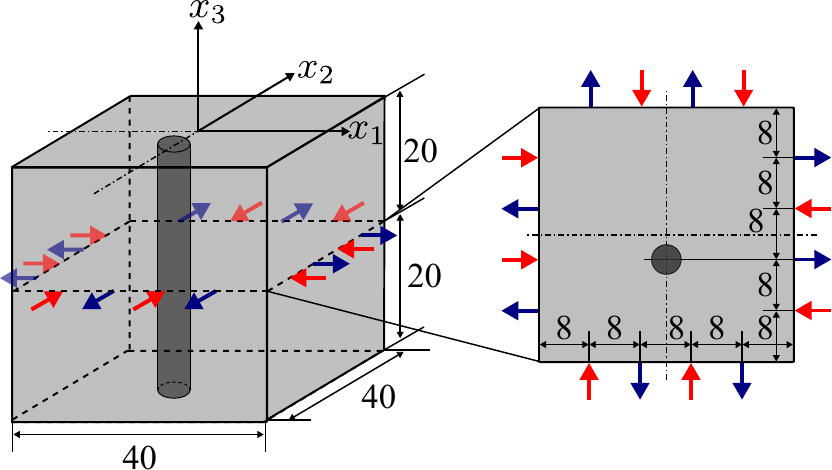}
\caption{
The numerical phantom. The unit of length is ${\rm mm}$. The absorber rod is shown in dark gray. Red and blue arrows show source fibers and detection fibers, respectively.
}
\label{solid:schem}
\end{figure}

Diffuse light in a cuboid-shaped numerical phantom is simulated by the finite element method implemented in TOAST \cite{Schweiger-Arridge14} and $\Phi^{(l)}$ ($l=1,2,\dots,M_{\rm SD}$) are computed (see below for the generation of the mesh). The numerical phantom, whose domain is denoted by $\Omega$, has a face of size $4\,{\rm cm}\times 4\,{\rm cm}$ and its height is $4\,{\rm cm}$. The absorption and reduced scattering coefficients are set to $\mu_{a,0}=0.02\,{\rm mm}^{-1}$ and $\mu_s'=0.85\,{\rm mm}^{-1}$. Here, the reduced scattering coefficient $\mu_s'$ is related to $D_0$ as $D_0=1/(3\mu_s')$. We put $\mathfrak{n}=1.52$ for the refractive index. In the numerical phantom, we placed an absorber rod of height $4\,{\rm cm}$. The center of the circle, which is the cross section of the rod, is at $(x_1,x_2)=(0,\,-4\,{\rm mm})$. This rod of diameter $5\,{\rm mm}$ has $\mu_a=0.06\,{\rm mm}^{-1}$ and $\mu_s'=0.85\,{\rm mm}^{-1}$. The rod has the same refractive index ($\mathfrak{n}=1.52$). We assume eight source fibers and eight detection fibers. They are attached to the numerical phantom at the height $2\,{\rm cm}$. See Fig.~\ref{solid:schem} for the numerical phantom and measurement setup. The reference data was obtained with a phantom which has the same optical properties but does not have the absorber rod.

When the diffusion equations (\ref{de0}) and (\ref{de1}) for $d=3$ were solved by the forward solver implemented in TOAST \cite{Schweiger-Arridge14}, almost uniform volume tetrahedral meshes were generated by Gmsh \cite{Geuzaine-Remacle09}. In this way, we compute $G(x,y)$ ($x\in\Omega$, $y\in\pp\Omega$) and obtain $\phi^{(l)}$ in (\ref{dataRytov}). The origin of the coordinates of the mesh is at the center of the top plane of the phantom ($-20\,{\rm mm}\le x_1\le 20\,{\rm mm}$, $-20\,{\rm mm}\le x_2\le 20\,{\rm mm}$, $-40\,{\rm mm}\le x_3\le 0\,{\rm mm}$). In the plane at $x_3=-20\,{\rm mm}$, positions $(x_1,x_2)$ of eight sources and eight detectors are given by 
$(-20,-4)$, $(-20,12)$, $(-12,-20)$, $(-4,20)$, $(4,-20)$, $(12,20)$, $(20,-12)$, $(20,4)$ for the sources and 
$(-20,-12)$, $(-20,4)$, $(-12,20)$, $(-4,-20)$, $(4,20)$, $(12,-20)$, $(20,-4)$, $(20,12)$ for the detectors.

In this three-dimensional case, $\Omega_{\rm square}$ and $\Omega_{\rm disk}$ are given by 
\be
\fl
\Omega_{\rm square}=
\left\{x\in\Omega;\;
\xi_1-\frac{\ell}{2}<x_1<\xi_1+\frac{\ell}{2},\;
\xi_2-\frac{\ell}{2}<x_2<\xi_2+\frac{\ell}{2},\;
-40\,{\rm mm}<x_3<0\right\},
\ee
\be
\fl
\Omega_{\rm disk}=
\left\{x\in\Omega;\;
(x_1-\xi_1)^2+(x_2-\xi_2)^2<\left(\frac{\ell}{2}\right)^2,\;
-40\,{\rm mm}<x_3<0\right\}.
\ee
That is, $\Omega_{\rm recon}$ is long in the $x_3$ direction and its cross section is a square or disk. We have
\be
N=4,
\ee
and
\be
\phi^{(l)}(S)\approx
\frac{\mu_{a,0}\eta_0(\Delta y)^3}{G(x_d^{(l)},x_s^{(l)})}
\sum_{y_i\in\Omega_{\rm recon}}G(x_d^{(l)},y_i)G(y_i,x_s^{(l)}),
\ee
where $\Delta y=1\,{\rm mm}$. We set
\ba
\xi_1&=
\frac{\xi_1^{\rm(max)}}{M}S_1,\quad-M\le S_1\le M,
\nonumber \\
\xi_2&=
\frac{\xi_2^{\rm(max)}}{M}S_2,\quad-M\le S_2\le M,
\nonumber \\
\ell&=
\frac{\ell^{\rm(max)}}{M}S_3,\quad 1\le S_3\le M.
\ea
Note that $f_{\rm prior}=0$ if $S_3$ is not positive or $S_4$ is less than $-m$,

\section{Results}
\label{results}

\subsection{Reconstruction in two dimensions}

\begin{figure}[ht]
\centering
\includegraphics[width=0.4\textwidth]{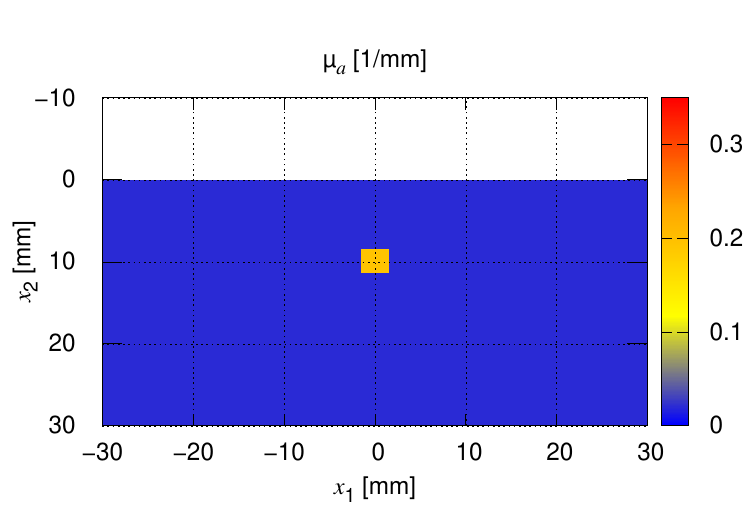}
\includegraphics[width=0.4\textwidth]{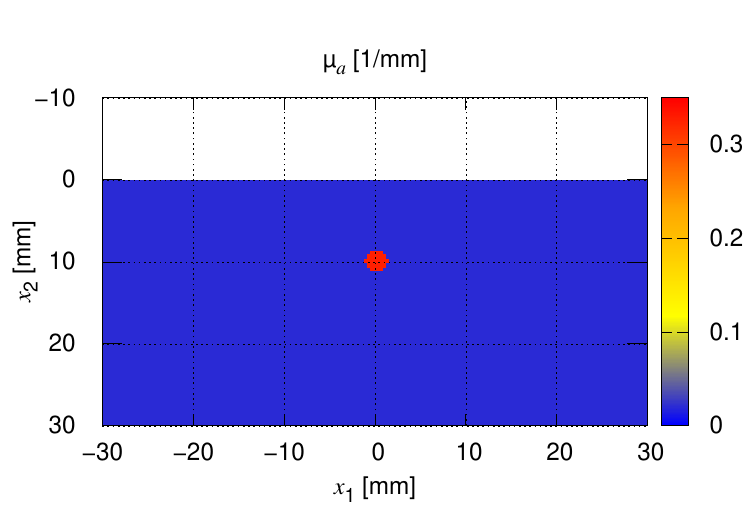}
\caption{
Reconstructed images for (Left) $\Omega_{\rm recon}=\Omega_{\rm square}$ and (Right) $\Omega_{\rm recon}=\Omega_{\rm disk}$.
}
\label{sa:fig1}
\end{figure}

Let us consider DOT described in Sec.~\ref{num2d}. Figures \ref{sa:fig1} shows reconstructed images. In Fig.~\ref{sa:fig1} (Left), $\Omega_{\rm recon}=\Omega_{\rm square}$. In Fig.~\ref{sa:fig1} (Right), $\Omega_{\rm recon}=\Omega_{\rm disk}$. The following parameter values were used.
\be
\varepsilon=10^{-5},\quad T_{\rm high}=10^{-4},\quad T_{\rm low}=10^{-14},\quad M=256.
\ee
We have $\xi_1^{\rm(max)}=4.5\,{\rm mm}$, $\xi_2^{\rm(max)}=3/64\,{\rm mm}$, $\ell^{\rm(max)}=9\,{\rm mm}$, and $\eta_0^{\rm(max)}=20$.

The initial guess was chosen as $\bar{S}=(0,M/2,1,0)$, that is initially,
\be
\xi_1=0,\quad\xi_2=0.047\,{\rm mm},\quad\ell=0.035\,{\rm mm},\quad\eta_0=0.
\ee
In the case of $\Omega_{\rm recon}=\Omega_{\rm square}$, the obtained configuration is $S_*=(-8,214,78,111)$, which reads
\be
\xi_1=-0.14\,{\rm mm},\quad\xi_2=10.03\,{\rm mm},\quad
\ell=2.74\,{\rm mm},\quad\mu_a=0.193.
\label{result2dsqu}
\ee
In the case of $\Omega_{\rm recon}=\Omega_{\rm disk}$, the obtained configuration is $S_*=(0,214,35,197)$, which reads
\be
\xi_1=0.00\,{\rm mm},\quad\xi_2=10.03\,{\rm mm},\quad
\ell=2.46\,{\rm mm},\quad\mu_a=0.328.
\label{result2dcir}
\ee

\subsection{Reconstruction in three dimensions}

\begin{figure}[ht]
\centering
\includegraphics[width=0.4\textwidth]{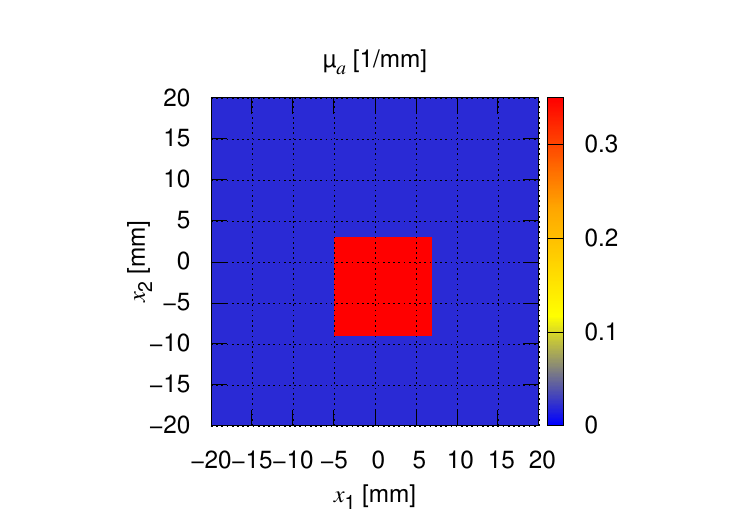}
\includegraphics[width=0.4\textwidth]{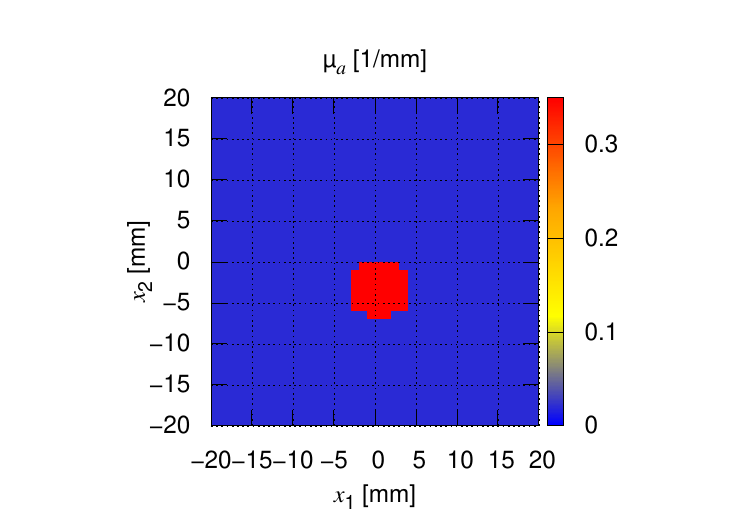}
\caption{
Reconstructed absorber rod for (Left) $\Omega_{\rm recon}=\Omega_{\rm square}$ and (Right) $\Omega_{\rm recon}=\Omega_{\rm disk}$. The true position of the center of the rod in the $x_1$-$x_2$ plane is $(0,\;-4\,{\rm mm})$.
}
\label{sa:fig2}
\end{figure}

Next we consider DOT which is described in Sec.~\ref{num3d}. The reconstructed absorber rod in the three-dimensional numerical phantom is shown in Fig.~\ref{sa:fig2}. In the left panel of Fig.~\ref{sa:fig2}, $\Omega_{\rm recon}=\Omega_{\rm square}$. In the right panel of Fig.~\ref{sa:fig2}, $\Omega_{\rm recon}=\Omega_{\rm disk}$. The parameters were chosen as follows.
\ba
&
M=256,\quad\varepsilon=10^{-5},\quad
\mu_{a,0}=0.02\,{\rm mm}^{-1},\quad
T_{\rm high}=1,\quad T_{\rm low}=10^{-13},
\nonumber \\
&
\xi_1^{\rm(max)}=10\,{\rm mm},\quad\xi_2^{\rm(max)}=10\,{\rm mm},
\quad\ell^{\rm(max)}=32\,{\rm mm},\quad\eta_0^{\rm(max)}=128.
\ea
We initially set $S_1=0$, $S_2=0$, $S_3=M/8$, $S_4=0$. That is, at first,
\be
(\xi_1,\xi_2)=(0,0),\quad
\ell=\frac{\ell^{\rm(max)}}{8}=4\,{\rm mm},\quad\eta_0=0.
\ee
The obtained values are
\be\fl
\cases{
\mbox{(square)}\quad
\xi_1=0.08\,{\rm mm},\quad\xi_2=-3.09\,{\rm mm},\quad
\ell=11.9\,{\rm mm},\quad\mu_a=0.37\,{\rm mm}^{-1},
\\
\mbox{(disk)}\quad
\xi_1=0.00\,{\rm mm},\quad\xi_2=-3.79\,{\rm mm},\quad
\ell=7.50\,{\rm mm},\quad\mu_a=1.14\,{\rm mm}^{-1}.
}
\label{result3d}
\ee

\section{Discussion}
\label{discussion}

In the case of the two-dimensional numerical phantom, reconstructed results in (\ref{result2dsqu}) and (\ref{result2dcir}) show that the position of the target $(\xi_1,\xi_2)$ is reconstructed more robustly than other parameters $\ell,\eta_0$. The same behavior is observed for the three-dimensional numerical phantom. The results in (\ref{result3d}) show that the reconstructed position $(\xi_1,\xi_2)$ is more accurate than the other parameters.

For the three-dimensional numerical phantom, compared with the position of the reconstructed target, the reconstructed $\mu_a$ in (\ref{result3d}) are not close to the true value of $\mu_a=0.06\,{\rm mm}^{-1}$. This attributes to the Rytov approximation. In Sec.~\ref{rytov}, higher-order terms in the Rytov series are ignored assuming $\eta$ is small. Since $\eta_0=2$ for $\mu_a=0.06\,{\rm mm}^{-1}$ and $\mu_{a,0}=0.02\,{\rm mm}^{-1}$ is not small, the reconstructed value of $\mu_a$ is not accurate.

If the inverse problem (\ref{dataRytov0}) is solved by the naive discretization of $y$, which is usually done, $N$ becomes the number of voxels in the medium plus $1$. On the other hand, $N=4$ in this paper. Thus the number of unknowns is significantly reduced by the constraint of the target shape. Moreover, since random numbers are used to move in the landscape of the cost function, our approach is not trapped by local minima. This is a significant superiority to conventional iterative methods such as the conjugate gradient method and Gauss-Newton method.

The $\ell^1$ norm is used for the regularization term in (\ref{SA:cost}). This is not the only choice. In our numerical scheme, different regularizations are possible as iterative schemes.

One natural next step is to extend the present numerical scheme to find multiple targets. If we have $n$ targets, the number of unknown parameters becomes $nN$. With another approach of the simulated annealing, we have shown that one thousand spins (i.e., the number of $S_i$ is $1000$) can be used to reconstruct the absorption coefficient of the diffusion equation \cite{Jiang-etal21}. Hence it is expected that the present method can be extended to reconstruct about one hundred targets ($n=100$).

\section{Conclusion}
\label{concl}

Through numerical experiments in two and three dimensions, we have shown that a target in the medium can be reconstructed by assuming a simple shape such as a square or a disk.

Simulated annealing is used for the Metropolis-Hastings algorithm to reach a converged result. In the numerical calculation, at first, different configurations of $S$ are tried. Eventually, only configurations which are close to each other are tested. Since this shift takes place slowly, configurations that are close to the true configuration are obtained. In this way, the target can be identified even when the initial guess is far from the true value. To obtain reconstructed values, $32760$ Monte Caro steps were necessary in the case of the two-dimensional numerical experiment, whose calculation takes about $100\,{\rm sec}$ on a laptop computer.

\section*{Acknowledgements}
YJ is supported by the National Natural Science Foundation of China (No.~11971121). NT is supported by JSPS KAKENHI (No.~JP16K05418). MM is supported by JSPS KAKENHI (No.~JP17K05572, JP18K03438).


\appendix

\section{Green's function in the half space}
\label{green}

Let us consider the half space in $\Rm^2$. In this case, the Green's function is obtained as
\be\fl
G(x,y)=\frac{1}{2\pi D_0}\int_0^{\infty}\frac{\cos(q(x_1-y_1))}{\lambda(q)}\left(e^{-\lambda(q)|x_2-y_2|}+\frac{\zeta D_0\lambda(q)-1}{\zeta D_0\lambda(q)+1}e^{-\lambda(q)(x_2+y_2)}\right)\,dq,
\ee
where
\be
\lambda(q)=\sqrt{\frac{\mu_{a,0}}{D_0}+q^2}.
\ee
The integral over $q$ can be evaluated by the double-exponential formula \cite{Ooura-Mori91}. Let ${\rm h}$ be a small number and $N_k$ be a large integer. Let us introduce
\be
q=\frac{\pi}{{\rm h}|x_1-y_1|}\phi(t),\quad
\phi(t)=\frac{t}{1-\exp(-6\sinh{t})},
\ee
and
\be
f(q)=\frac{1}{\lambda(q)}\left(e^{-\lambda(q)|x_2-y_2|}+\frac{\zeta D_0\lambda(q)-1}{\zeta D_0\lambda(q)+1}e^{-\lambda(q)(x_2+y_2)}\right)\cos\left(q(x_1-y_1)\right).
\ee
Then we have
\be
G(x,y)\approx\frac{1}{2D_0|x_1-y_1|}\sum_{k=-N_k}^{N_k}f\left(\frac{\pi}{{\rm h}|x_1-y_1|}\phi(k{\rm h}+\frac{{\rm h}}{2})\right)\phi'(k{\rm h}+\frac{{\rm h}}{2}).
\ee


\end{document}